\documentclass{webofc}
\usepackage[varg]{txfonts}   
\usepackage{hyperref}
\usepackage[mathlines]{lineno}
\begin{document}
\title{Tuning the CMS Coffea-casa facility for 200 Gbps Challenge}

\author{\firstname{Sam} \lastname{Albin}\inst{1}
\and
\firstname{Garhan} \lastname{Attebury}\inst{1}
\and
\firstname{Kenneth} \lastname{Bloom}\inst{1}
\and
\firstname{Brian} \lastname{Bockelman}\inst{2}
\and
\firstname{Benjamin Tovar} \lastname{Lopez}\inst{3}
\and
\firstname{Carl} \lastname{Lundstedt}\inst{1}
\and
\firstname{Oksana} \lastname{Shadura}\inst{1}
\fnsep\thanks{\email{oksana.shadura@cern.ch}}
\and
\firstname{John} \lastname{Thiltges}\inst{1}
\and
\firstname{Derek} \lastname{Weitzel}\inst{1}
\and
\firstname{Andrew} \lastname{Wightman}\inst{1}
}

\institute{
University of Nebraska-Lincoln, Lincoln, NE 68588 
\and
Morgridge Institute for Research, Madison, WI 53715
\and
University of Notre Dame, Notre Dame, IN 46556,
}

\abstract{%
As a part of the IRIS-HEP “Analysis Grand Challenge” activities, the Coffea-casa AF team executed a “200 Gbps Challenge”. One of the goals of this challenge was to provide a setup for execution of a test notebook-style analysis on the facility that could process a 200 TB CMS NanoAOD dataset in 20 minutes.

We describe the solutions we deployed at the facility to execute the challenge tasks. The facility was configured to provide 2000+ cores for quick turn-around, low-latency analysis. To reach the highest event processing rates we tested different scaling backends, both scaling over HTCondor and Kubernetes resources and using Dask and Taskvine schedulers. This configuration also allowed us to compare two different services for managing Dask clusters, Dask labextention, and Dask Gateway server, under extreme conditions.

A robust set of XCache servers with a redirector were deployed in Kubernetes to cache the dataset to minimize wide-area network traffic. The XCache servers were backed with solid-state NVME drives deployed within the Kubernetes cluster nodes. All data access was authenticated using scitokens and was transparent to the user.
To ensure we could track and measure data throughput precisely, we used our existing Prometheus monitoring stack to monitor the XCache pod throughput on the Kubernetes network layer. Using the rate query across all of the 8 XCache pods we were able to view a stacked cumulative graph of the total throughput for each XCache. This monitoring setup allowed us to ensure uniform data rates across all nodes while verifying we had reached the 200 Gbps benchmark.
}
\maketitle
\section{Introduction: Analysis Grand Challenge as first step for 200 Gbps challenge}

The IRIS-HEP Analysis Grand Challenge (AGC)~\cite{held_alexander_2022} began as an integration initiative for IRIS-HEP, aimed at developing methods for testing a complete analysis pipeline. Its purpose is to prepare infrastructure for the HL-LHC, focusing on a realistic-scale physics analysis, and to create flexible, user-friendly, low-latency analysis pipeline that could be used by analysts. Additionally, it offers an opportunity to assess the new Python ecosystem for analysis.

The AGC consists of two main components: defining a physics analysis task that reflects HL-LHC requirements and implementing an analysis pipeline to address this task. The AGC provides a clearly defined physics analysis task along with a pipeline implementation, enabling scientists to identify and resolve performance bottlenecks and usability issues. Several AGC implementations have been developed using the Coffea framework\cite{coffea}, RDataFrame\cite{rdataframe}, Julia\cite{julia}, and ColumnFlow\cite{columnflow}.

The AGC aims to bridge the gap in large-scale analysis capabilities for the HL-LHC and strengthen connections to LHC experiments.

\section{Preparing next generation of Analysis Facilities for HL-LHC scale: the 200 Gbps setup}

A project that brought together many people from within IRIS-HEP and beyond was the “200 Gbps challenge”. While there is currently limited agreement in the broader field on exactly what a “representative” physics analysis in the future may look like, the ability to reach significant data rates in physics analysis in order to achieve “interactive” turnaround and speed up time-to-insight for physicists is an important one. We converged on a concrete target for a data rate to demonstrate: 200 Gbps, corresponding to reading 25\% of 180 TB of data over half an hour. In order to be as relevant to the reality of ATLAS and CMS as possible, this challenge ran as two projects: one using ATLAS data in PHYSLITE format \cite{physlite}, one using CMS data in NanoAOD format \cite{nanoaod}. Both projects were closely aligned with many common discussions and solutions to problems faced.

In order to reach this ambitious goal for data rates, we employed the University of Chicago and University of Nebraska-Lincoln analysis facilities in setups optimized for throughput. The challenge however was not only streaming data over the network, but also making it available for physics. In practice this meant data being read out of XCaches\cite{bauerdick2014xrootd}, read, decompressed, and made available as arrays in memory, ready for any further analysis work (which strongly depends on the physics analysis use case). We employed a variety of setups throughout this challenge, starting with code developed in previous AGC work and extending it further to feature additional benchmarking utilities.

The goal of reading 25\% of 180 TB of data over half an hour would require reading the 45~TB of data into the analysis process. We assume we read 25\% of the data from the NanoAOD, meaning 180~TB of NanoAOD is required to push 45~TB of branches. At 2~KB/event (the expected CMS NanoAOD data format size for Run-3), 180~TB of NanoAOD would consist of 96 billion events. Reading this many events in 30 minutes requires a sustained 55~MHz event rate. Our sample AGC analysis could run at 25~KHz per core, meaning 2,200 cores are needed to sustain the 55~MHz event rate.

This contribution focuses on CMS relevant setup and  describes the analysis facility components that were deployed and tested during the exercise: XCache caching service, enabled diverse computing executors available in the analysis pipeline, the hardware setup that provided the computing resources for 200 Gbps challenge, and monitoring utilities which were used over the tests.

\subsection{XCache setup}

The XRootD software framework\cite{rootd} is essential to various LHC initiatives. The Coffea-Casa AF includes its integrated ``XCache'' configuration for CMS experimental datasets, enabling faster data access during interactive analysis, particularly when users need to repeatedly access the same dataset. The main functionality offers data caching service mostly for analysis workflows executed at the facility. To support high-throughput caching, NVMe storage devices from an existing Ceph pool were repurposed as RAID0 arrays across each of the eight nodes. The XCache pods were allocated 64 GiB of RAM and six dedicated CPU cores to ensure high I/O performance. Local storage persistent volumes (PVs) were statically provisioned for each XCache deployment. Additionally, each XCache instance was assigned a unique hostname and secured with an SSL certificate to ensure accessibility.

\subsection{Providing the diverse computing executors for 200 Gbps challenge}

A central design principle of the Coffea-Casa AF is to make use of existing compute and storage infrastructure to support analysis tasks. For Coffea-Casa to be widely adopted, it needs to leverage existing experiment infrastructure without requiring changes to the existing setup.
To meet this design goal, Coffea-Casa developed  seamless integration with existing resources, including an HTCondor \cite{condor-practice} batch cluster and a Kubernetes cluster, both hosted at Nebraska.

Over last years we have explored dynamic task scheduling optimized for interactive computations, such as Dask, a popular parallel computing library\cite{dask}. One of Dask's notable features is its ability to be easily deployed on traditional batch queuing systems like PBS, Slurm, LSF, and HTCondor through Dask-jobqueue \cite{dask-jq}. Given the modularity and extensive functionality of Dask, as well as the need to maintain customizations for different analysis facility environments,   a high-level extension of the Dask-jobqueue module and integration with Dask-Gateway was developed, allowing the deployment of a multi-tenant server for managing Dask clusters.

In the case of Dask and HTCondor setup, when the Coffea-Casa AF workload grows, it submits Dask jobs to the Nebraska HTCondor queue. These jobs launch containers with Dask worker services that connect to the Dask scheduler created when the user logged in. This allows HTCondor resources to be easily used for Dask tasks on behalf of the Coffea-Casa user. We explored this setup as a baseline test for 200 Gbps measurements at UNL.

The modular architecture of Coffea-Casa AF also makes it simple to integrate with other types of batch resources as well using other task scheduling frameworks such as TaskVine\cite{Taskvine}. The TaskVine library is particularly useful in scenarios requiring the management of a large number of tasks that can run concurrently, such as in scientific computing, data processing, or machine learning workflows. It supports dynamic scheduling, which is important for interactive computations where tasks need to be scheduled in real time based on their dependencies and available resources.

TaskVine's capabilities are designed to work well with distributed systems, enabling easy integration with various task queues, job schedulers, and parallel computing frameworks. In addition TaskVine has a dedicated Dask executor that allows the TaskVine library to work seamlessly with the Dask distributed computing framework. In this contribution we explored as well a setup where we integrated TaskVine in our workflow. In this case TaskVine’s executor was managing the submission of tasks to Dask through Dask scheduler and scale to available HTCondor workers and ensures that tasks are properly distributed and executed based on their dependencies.

In addition, members of the Nebraska team tested Coffea-Casa AF setup utilizing Dask Gateway \cite{dask-gateway} which allowed efficient use of the available Kubernetes resources while improving the management of multi-tenant Dask cluster used for 200 Gbps. This third setup achieved the targeted 200 Gbps with significantly smaller number of workers.

\subsection{Flexible computing resource provisioning models for 200 Gbps challenge}

\begin{figure}[h]
\centering
\includegraphics[width= 0.8\textwidth]{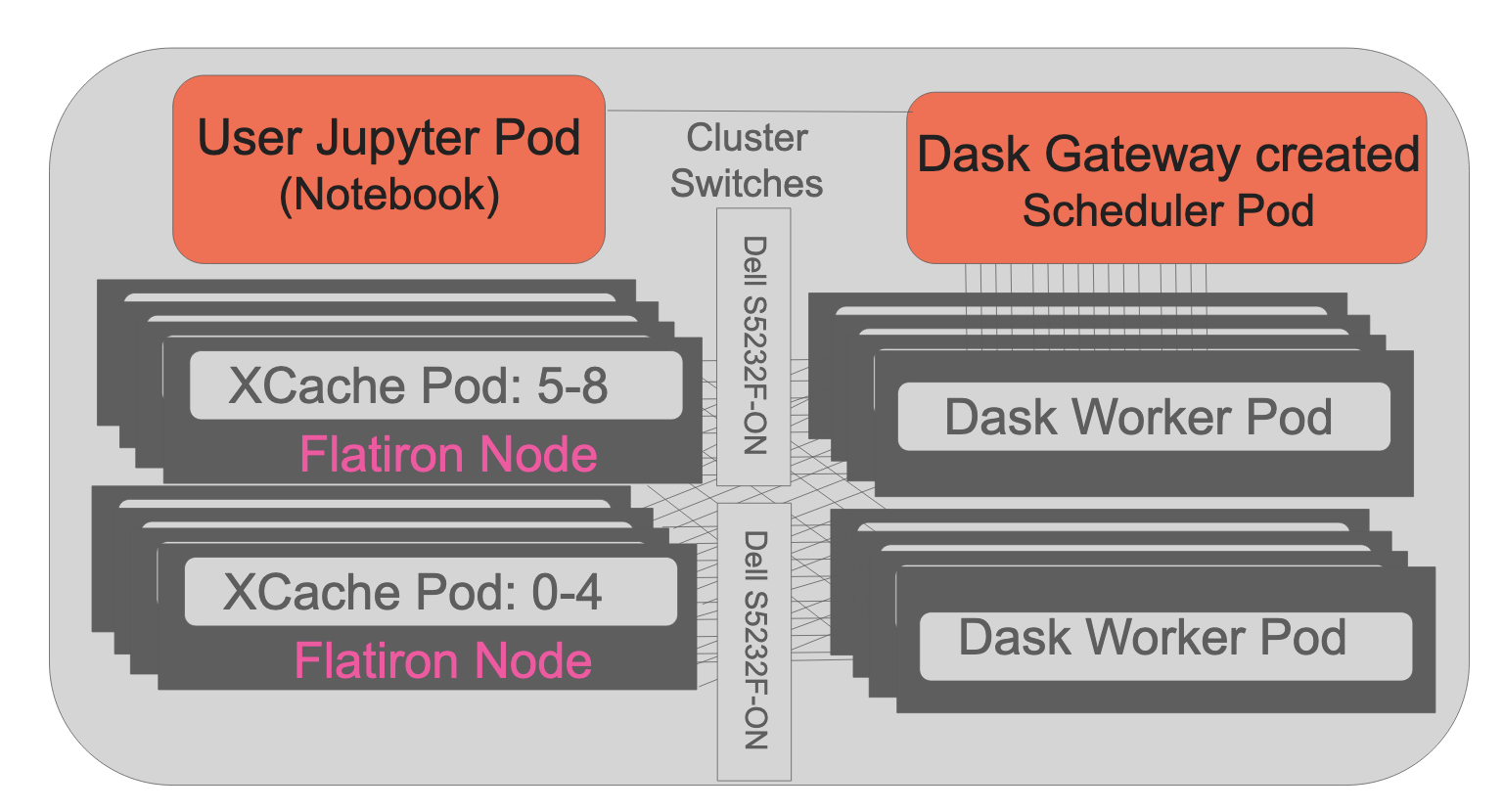}
\caption{Flexible computing resources utilizing Kubernetes workers}
\label{fig-daskgateway}
\end{figure}

\begin{figure}[h]
\centering
\includegraphics[width= 0.8\textwidth]{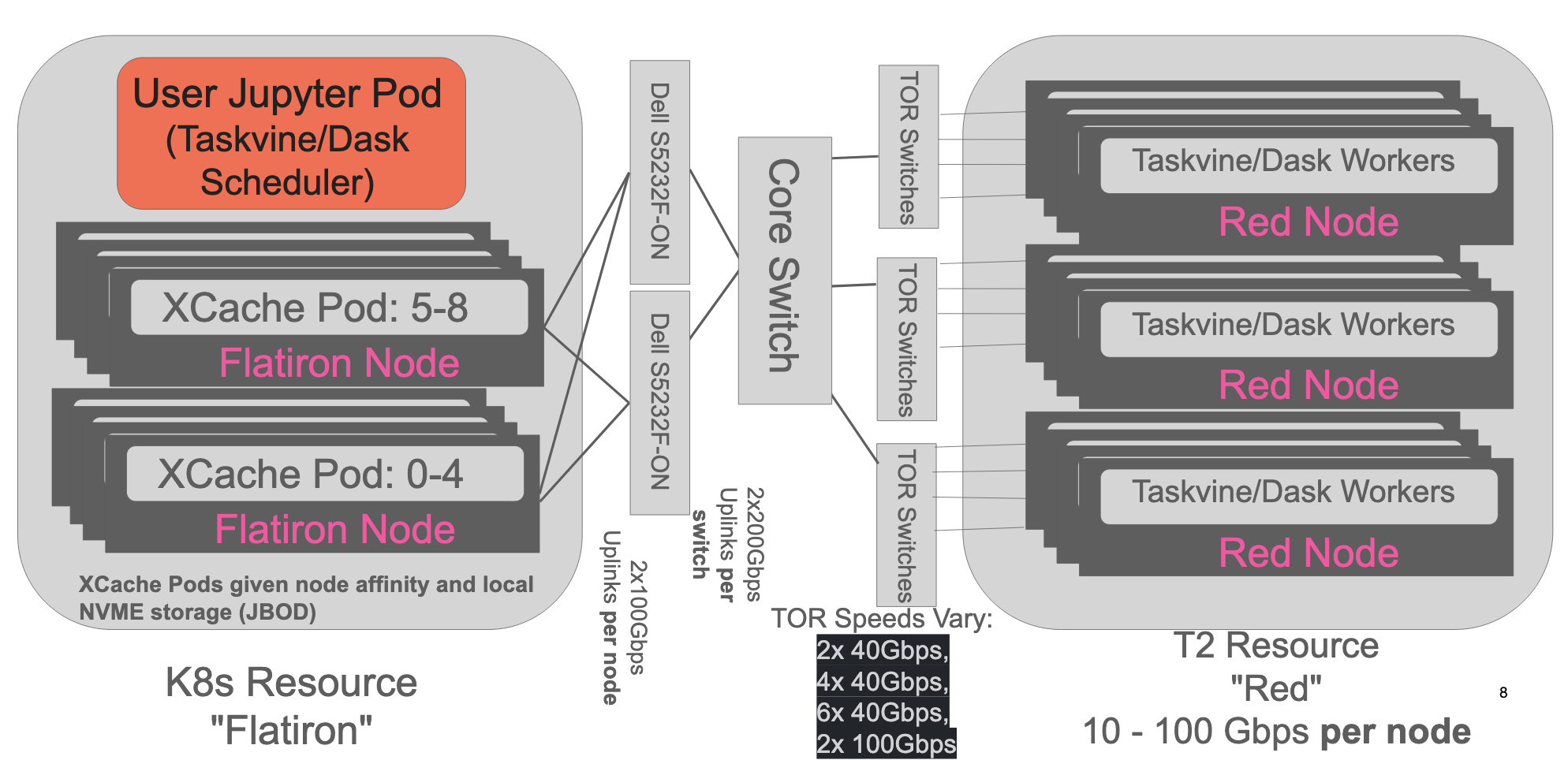}
\caption{Flexible computing resource utilizing external HTCondor workers}
\label{fig-condor}
\end{figure}

Although the scalability and parallelization of the data analysis is handled by the Dask software library, the underlying hardware providing the actual computational resource was configured using two different topologies.  The first method in which computational resources were deployed utilized Dask Gateway and Dask pods running natively in the Kubernetes resource called Flatiron, see figure \ref{fig-daskgateway}.  The second resource deployment involved Taskvine\cite{Taskvine} allocating HTCondor workers in a general use HTCondor cluster that was external to the Kubernetes resource Flatiron, see figure \ref{fig-condor}.  

The first cluster type, utilizing solely Kubernetes resources, ensured that all network traffic between XCache and the worker nodes was confined to the Kubernetes resource, allowing large bandwidth and little latency.  However, this resource was constrained by the limited number of CPUs available in the Flatiron cluster, around 500.  The second cluster type, utilizing HTCondor, demanded that data traverse physical network boundaries to get to the analyzing CPU but was not nearly as restricted in the number of CPUs available for the analysis due to the larger CPU count, over 2000 dedicated cores, available in the HTCondor cluster called Red.

\subsection{User monitoring as a part of 200 Gbps challenge}

The crucial measurements of bandwidth for this experiment were performed using Prometheus\cite{prometheus}. By applying the Prometheus ``irate'' query to the network interfaces of the eight XCache pods in the Flatiron Kubernetes cluster, Prometheus provided real-time insights into data transmission rates for each service pod. To reduce network overhead, the `externalTrafficPolicy` of the XCache services was set to `Local`. This monitoring setup played a crucial role in verifying that the 200 Gbps benchmark was successfully achieved.

\begin{figure}[h]
\centering
\includegraphics[width= \textwidth]{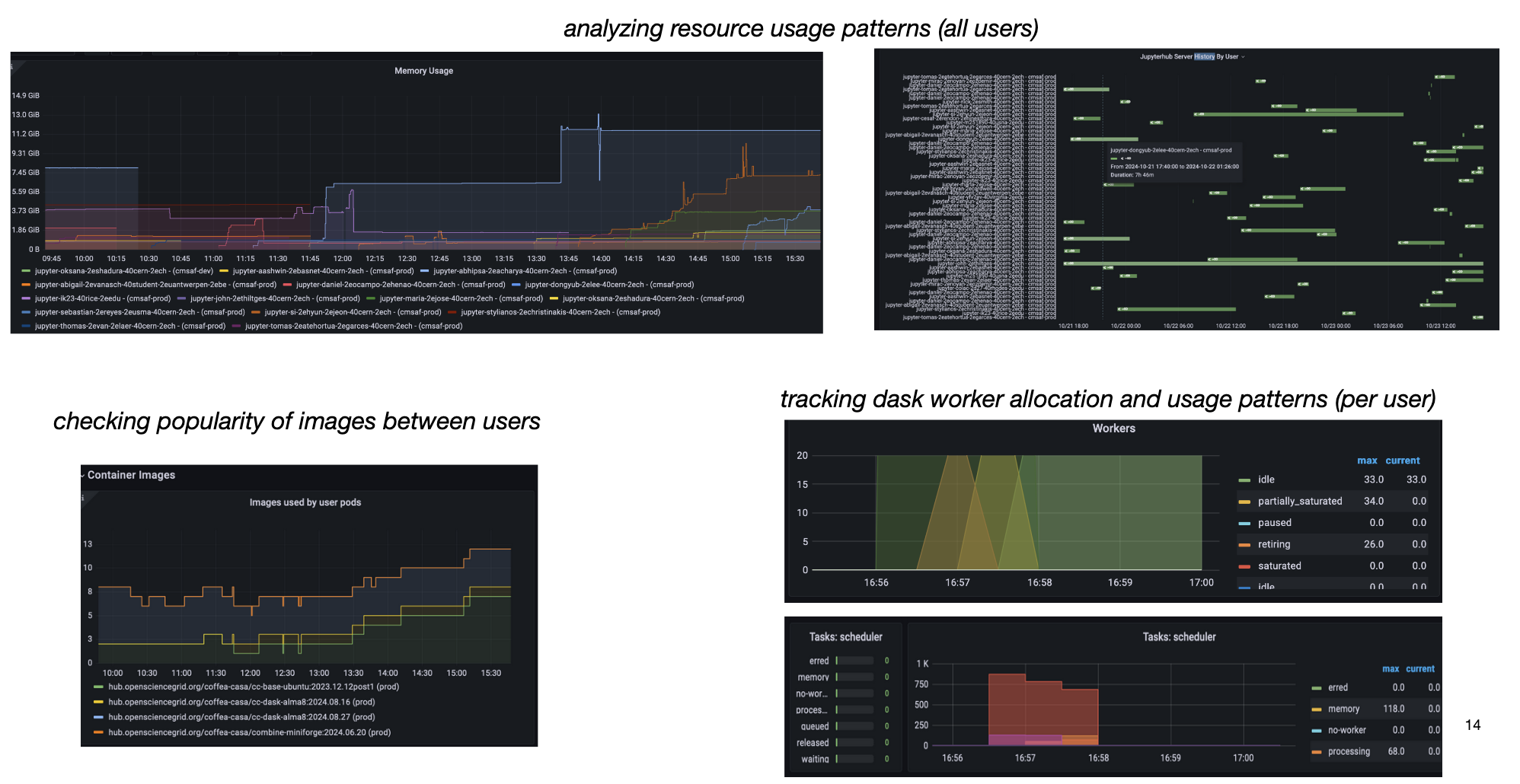}
\caption{Monitoring}
\label{fig-2}
\end{figure}

\section{200 Gbps measurements in the different deployments}

\begin{figure}[h]
\centering
\includegraphics[width=\textwidth]{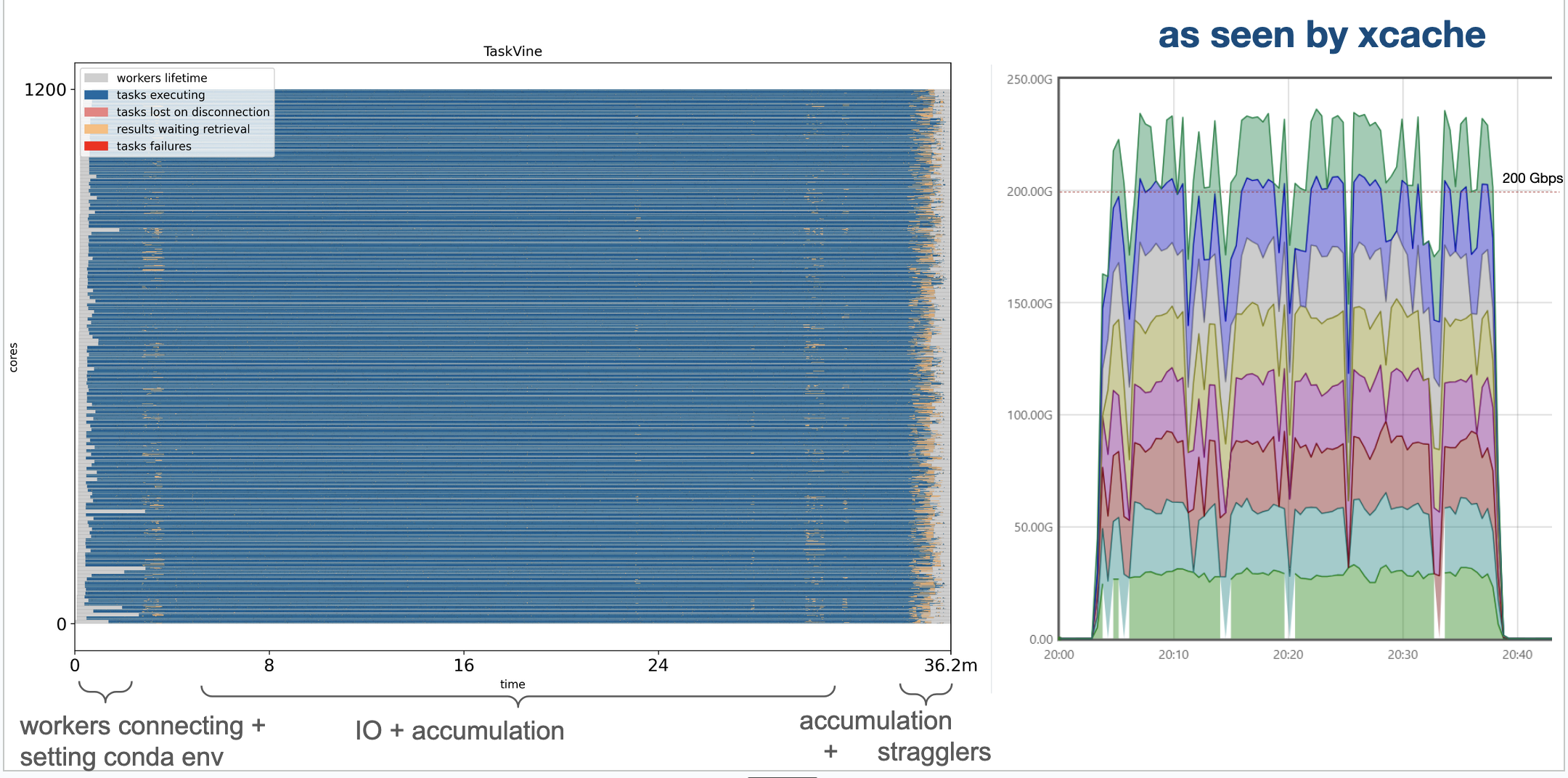}
\caption{200 Gbps measurements in TaskVine cluster setup}
\label{fig-3}
\end{figure}

Figure \ref{fig-3} shows measurements with the TaskVine setup executed over 1200 HTCondor workers. The task graph shows smooth execution over stages from the setup of workers, handling I/O and accumulation specific tasks. The XCache monitoring showed  that with TaskVine we managed to achieve more then 200 Gbps over the execution time of a bit less then 36 minutes. 

\begin{figure}[h]
\centering
\includegraphics[width=0.5\textwidth]{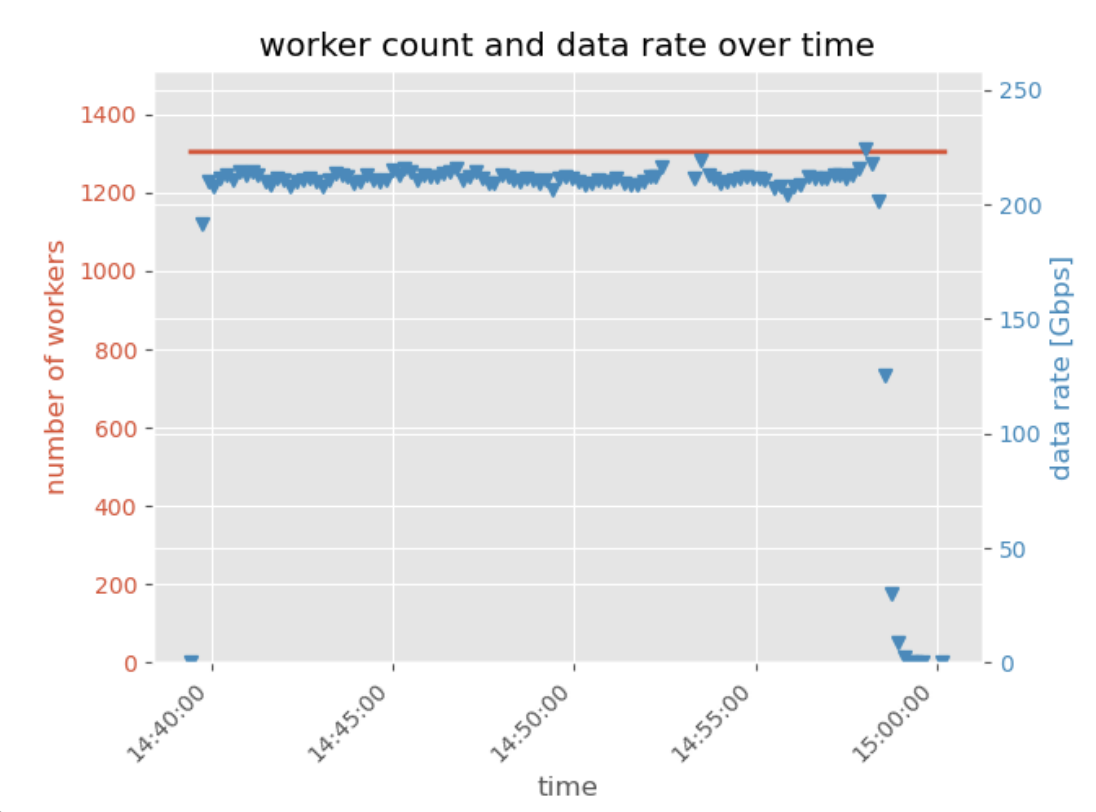}
\caption{200 Gbps measurements in HTCondor cluster setup}
\label{fig-4}
\end{figure}

Figure \ref{fig-4} shows measurements with the Dask and HTCondor setup executed over more then 1200 HTCondor workers successfully delivering 200 Gbps target data rate. 

\begin{figure}[h]
\centering
\includegraphics[width=\textwidth]{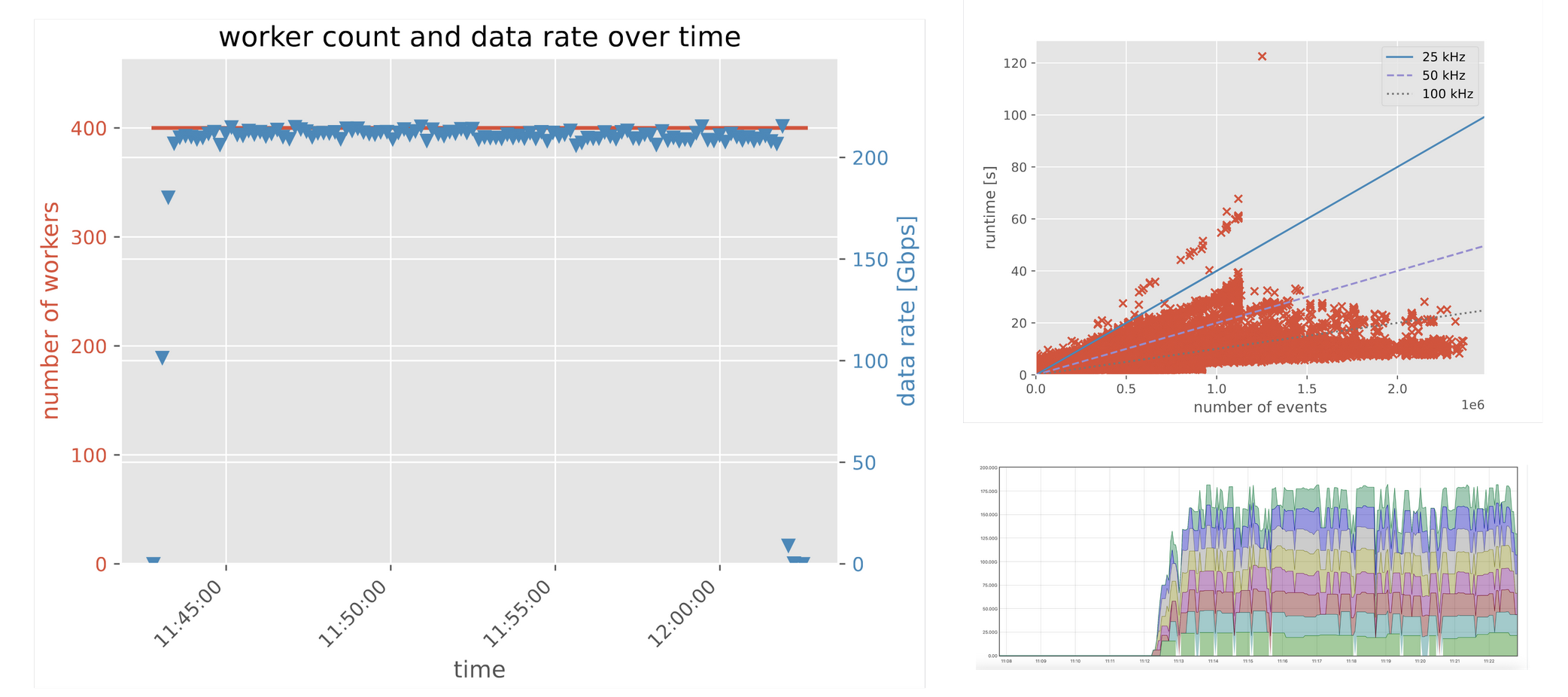}
\caption{200 Gbps measurements in Kubernetes cluster setup}
\label{fig-5}
\end{figure}

Figure \ref{fig-5} shows measurements with Dask-Gateway over Kubernetes cluster setup, where we were able to achieve running at the 200 Gbps target, executing workflow over more then 400 Kubernetes workers. We were also able to monitor the runtime needed for each of the files to be processed in our dataset as seen by XCache and on average we were able to achieve 100 kHz event ratio as a result of well-configured XCache, and good network layout and hardware.

\section{Lessons learned}

As a result, we consider the 200 Gbps a very successful exercise format that stimulated a huge amount of progress and activity within 8 weeks together with the whole team. We highlight these lessons learned:

\begin{itemize}
\item In the  setup described in this contribution, we realized we still have room for improvement,  since we faced some challenges over the execution with memory use and scaling to all available resources.
\item The NanoAOD format compression algorithm had a very large effect on performance of 200 Gbps workflow: switching from LZMA to ZSTD compression algorithms brought a 2.5x event processing rate improvement.
\item Scaling Dask to 2000+ workers generally worked fine, but in the future  more time would be needed for testing combining large numbers of workers and very complex graphs.
\item Good performance was also observed with TaskVine as alternative scheduler for graphs.
\end{itemize}

\section{Conclusions}

Coffea-Casa is an analysis facility focused on integrating the various technologies from IRIS-HEP and CMS experiment. The main goal of Coffea-casa project is to make new developments available to early users in CMS and ATLAS experiments. The scale of 200 Gbps challenge allowed to identify new bottlenecks in the analysis tools and in facility setup (many of which have already been fixed) and  to better understand  possible limitations for end-user analysis before HL-LHC, especially focusing on the analysis throughput.

The next goal for the Nebraska Coffea-casa deployment is to focus on trying to better understand  the memory issues with large graphs shown during exercise and try to execute the next challenge where we are going to try to achieve 400 Gbps in a similar setup.

\section{Acknowledgements}
This work was supported by the National Science Foundation under Cooperative Agreements OAC-1836650, PHY-2121686 and PHY-2323298.

\bibliography{bib/template}

\end{document}